\documentclass[twocolumn,final]{revtex4}
\usepackage{amscd}
\usepackage{amsmath}
\usepackage{amssymb}
\usepackage{amsthm}
\usepackage{graphicx}
\usepackage{mathrsfs}
\usepackage{hyperref}
\usepackage{mathptmx}
\usepackage{subfig}
\usepackage{color}
\usepackage[usenames,dvipsnames]{xcolor}
\usepackage{enumerate}
\usepackage{comment}
\newcommand{\al}{\alpha}

\newcommand{\de}{\delta}
\newcommand{\De}{\Delta}
\newcommand{\ep}{\epsilon}
\newcommand{\fr}{\frac}

\newcommand{\lf}{\left}

\newcommand{\om}{\omega}

\newcommand{\rg}{\right}

\newcommand{\Si}{\Sigma}

\newcommand{\scr}{\mathscr}

\begin{document}
\title{Global Level Number Variance in Integrable Systems}
\author{Tao Ma, R.A. Serota}
\email{serota@ucmail.uc.edu}
\affiliation{Department of Physics\\
University of Cincinnati\\
Cincinnati, OH 45244-0011}
\date{\today}
\begin{abstract}
We study previously un-researched second order statistics -- correlation function of spectral staircase and global level number variance -- in generic integrable systems with no extra degeneracies. We show that the global level number variance oscillates persistently around the saturation spectral rigidity. Unlike other second order statistics -- including  correlation function of spectral staircase -- which are calculated over energy scales much smaller than the running spectral energy, these oscillations cannot be explained within the diagonal approximation framework of the periodic orbit theory. We give detailed numerical illustration of our results using four integrable systems: rectangular billiard, modified Kepler problem, circular billiard and elliptic billiard.
\end{abstract}

\maketitle

\section{Introduction}

Interest in semiclassical properties of classically integrable systems picked up recently with deeper understanding of large persistent oscillations of the level number variance over an energy interval as a function of the interval width and of the phenomenon of level repulsion as manifested through deviations from the Poisson statistics of nearest level spacings. \cite{wickramasinghe05, wickramasinghe08, ma10mk, ma11eb, ma10lr} The precise nature of these effects are revealed by the structure of the correlation function of the level density and all relevant quantities can be computed both from the periodic orbit (PO) theory \cite{berry1985} and by direct quantum-mechanical calculation \cite{ma10mk}.

An early attempt of evaluation of the correlation function of spectral staircase (SS) \cite{serota08} and interest in global level number variance (GV) were motivated by the fluctuations of thermodynamic quantities of mesoscopic electronic systems \cite{serota09}. For instance, evaluation of orbital magnetic response in the integrable circumstance \cite{vonoppen1994,richter1996} calls for performing ensemble averaging (achieved via parametric averaging \cite{wickramasinghe05,wickramasinghe08,ma10mk}) prior to thermal averaging, which, in turn, requires knowledge of the magnetic field dependence of the correlation function of SS.

The central result of this work is to establish, theoretically and numerically, that GV exhibits large persistent oscillations around the saturation spectral rigidity. \cite{wickramasinghe05,wickramasinghe08,ma10mk,ma11eb,berry1985} Moreover, it is shown that these oscillations cannot be described in the standard framework that utilizes the diagonal approximation (DA) of the PO theory \cite{berry1985} but rather require an account of interference between periodic orbits with different winding numbers. Additionally, we evaluate the correlation function of SS and show that it can be expressed in terms of interval level number variance. 

This paper is organized as follows. In Sec. II, we evaluate GV and the correlation function of SS using the PO theory. In Sec. III, we present numerical evaluation of the GV vis-a-vis the saturation spectral rigidity for the rectangular, circular and elliptic billiards (RB, CB, EB) and the modified Kepler problem (MK). For RB, we proceed with a more extended analysis of the SS, its correlation function and interference effects in GV.

\section{Theory}

\subsection{Correlation Function of Spectral Staircase}

In PO theory, the fluctuating part of SS, $\de\scr{N}(\ep) \equiv \scr{N}(\ep) - \langle \scr{N}(\ep) \rangle$, is found as a sum over POs and their time-reversals \cite{berry1985}:
\begin{equation}\label{eq:deN}
\de \scr{N}(\ep) = \fr{2}{\hbar^{\mu}} \sum_j \de_j \fr{A_j(\ep)}{T_j(\ep)}  \sin(S_j(\ep)/\hbar),  \\
\end{equation}
where $\mu=(N-1)/2$ with $N$ the dimensionality of the position space \cite{berry1985} and $\de_j = 1/2$ if the PO and its time-reversal coincide and 1 otherwise. Fluctuations of SS and fluctuations of level density $\de\rho(\ep)$ are related via \cite{berry1985}
\begin{equation}\label{eq:derho}
\fr{\partial\de \scr{N}(\ep)}{\partial\ep} = \de\rho(\ep)  = \fr{2}{\hbar^{\mu+1}} \sum_j \de_j A_j (\ep) \cos\fr{S_j(\ep)}{\hbar} ,
\end{equation}
with the use of $T_j = dS_j/d\ep$ and on the account of the fact that the dominant contribution comes from differentiation of the oscillating term.

The correlation function of SS is found from (\ref{eq:deN}) as
\begin{equation}\label{eq:deNdeN}
\begin{split}
K_{\scr{N}}(\ep, \om) &\equiv \langle \de \scr{N}(\ep_1) \de \scr{N}(\ep_2) \rangle \\
&=
\fr{2}{\hbar^{2\mu}} \sum_j\de_j^2\fr{A_j^2(\ep)}{T_j^2(\ep)}
\left[  \cos\fr{\om T_j(\ep)}{\hbar} -  \cos\fr{2S_j(\ep)}{\hbar}  \right] \\
&\approx
\fr{2}{\hbar^{2\mu}} \sum_j\de_j^2 \fr{A_j^2(\ep)}{T_j^2(\ep)}
 \cos\fr{\om T_j(\ep)}{\hbar} ,
\end{split}
\end{equation}
where $\ep = (\ep_1 + \ep_2)/2$ and $\om = \ep_2 - \ep_1 \ll \ep$. For integrable systems, ensemble averaging is understood as the parametric averaging \cite{wickramasinghe05,wickramasinghe08,ma10mk} and the second, rapidly oscillating cosine was dropped in (\ref{eq:deNdeN}) as it produces a negligible contribution upon such averaging insofar as $\om$-dependence is concerned.

In what follows, unless explicitly stated otherwise, we drop the argument of $A_j$ and $T_j$. We notice that the (interval) level number variance over the energy interval of width $\om$ is given by \cite{wickramasinghe08}
\begin{equation}\label{eq:Sigma}
\Si(\ep,\om)
= \fr{4}{\hbar^{2\mu}} \sum_j \de_j^2 \fr{A_j^2 }{T_j^2} \left(1 - \cos\fr{\om T_j}{\hbar} \right) .
\end{equation}
Consequently, combining (\ref{eq:deNdeN}) and (\ref{eq:Sigma}), we have \cite{serota08}
\begin{equation}\label{eq:Kdiag}
\begin{split}
\ K_{\scr{N}}(\ep, \om) &= \De_3^\infty(\ep) - \fr{1}{2}\Si(\ep,\om) \\
&\approx \De_3^\infty(\ep) - \fr{\left|\om\right|}{2\De},  \,\,\,\,\,\, \left|\om\right| \ll \sqrt{\ep \De}
\end{split}
\end{equation}
where $\De$ is the mean level spacing and $\De_3^\infty(\ep)$ is the saturation spectral rigidity given by \cite{wickramasinghe08}
\begin{equation}\label{eq:Delta3}
\De_3^\infty(\ep) = \fr{2}{\hbar^{2\mu}} \sum_j \de_j^2 \fr{A_j^2 }{T_j^2} .
\end{equation}
Below, it is ordinarily assumed that $\om\ge0$. As expected, differentiation on $\ep_{1,2}$ inside the cosine of the last equation of (\ref{eq:deNdeN}), also gives the correlation function of the level density \cite{wickramasinghe08}, which can also be obtained directly from (\ref{eq:derho}).

\subsection{Global Level Number Variance}

We now turn to GV of SS, which is defined as follows: \cite{serota08}
\begin{equation}\label{eq:global_variance}
\Si_g( \ep )
\equiv \langle [\de\scr{N}(\ep) ]^2 \rangle
\equiv \langle [\scr{N}(\ep) - \langle\scr{N}(\ep)\rangle ]^2 \rangle .
\end{equation}
Formally, GV is a particular case of $\ K_{\scr{N}}(\ep, \om)$ in (\ref{eq:deNdeN}) with $\om=0$. We point out, however, that (\ref{eq:deNdeN}) was obtained using DA, which is sufficient for evaluation of the $\om$-dependence of $\ K_{\scr{N}}(\ep, \om)$ for a given $\ep$ (see supporting numerical evidence below). \footnote{Evaluation of correlation functions at small $\om$ presents difficulty similar to computations outside the DA framework, namely, a necessity to consider small differences between large actions;  in either circumstance, the result is not perturbative in nature.} However, it breaks down when the $\ep$-dependence of $\Si_g( \ep )$ is considered. Whereas DA yields, upon averaging,
\begin{equation}\label{eq:global_variance2}
\Si_g(\ep)
= \De_3^\infty(\ep)
-\fr{2}{\hbar^{2\mu}} \sum_j \fr{A_j^2 \de_j^2}{T_j^2} \cos\fr{2S_j}{\hbar}  \approx \De_3^\infty(\ep),
\end{equation}
interference between POs in accordance with (\ref{eq:deN})
\begin{equation}\label{eq:global_variance_full}
\begin{split}
\Si_g(\ep)
&= \fr{4}{\hbar^{2\mu}} \lf(\sum_j \fr{A_j \de_j}{T_j} \sin\fr{S_j}{\hbar} \rg)^2 ,
\end{split}
\end{equation}
must be considered to account for the full $\ep$-dependence of  $\Si_g(\ep)$ obtained in the numerical calculation below -- namely, the persistent oscillations of $\Si_g(\ep)$ around $\De_3^\infty(\ep)$.

Indeed, the off-diagonal contribution to GV contains, per (\ref{eq:global_variance_full}), the following term:
\begin{equation}\label{eq:global_variance_off-diag}
\begin{split}
&\sin\fr{S_j(\ep,\al)}{\hbar} \sin\fr{S_i(\ep,\al)}{\hbar}=  \\
&\fr{1}{2}\left[\cos\fr{S_j(\ep,\al)-S_i(\ep,\al)}{\hbar} - \cos\fr{S_j(\ep,\al)+S_i(\ep,\al)}{\hbar}\right]
\end{split}
\end{equation}
where the dependence of the action on the system parameter $\al$ is explicitly indicated. Parametric averaging involves integration over the distribution function $\rho(\al)$, such as a Gaussian distribution centered around the central value $\al_0$. \cite{ma10mk} Ordinarily, such integration with rapidly oscillating terms in (\ref{eq:global_variance_off-diag}) produces negligible contributions (as is the case with the dropped term in (\ref{eq:deNdeN}) and (\ref{eq:global_variance2}) as well as with the off-diagonal contribution when $\om \ne 0$). The notable exception occurs when $\partial_\al (S_j(\ep,\al) \pm S_i(\ep,\al))|_{\al=\al_0}=0$ and the arguments of the cosines scale as $(\al-\al_0)^2$. Below we illustrate this circumstance on RB.

\subsubsection{Rectangular Billiard}

For a particle of mass $m$ in a RB with sides $a$ and $b$, the amplitude, period and action of a PO with winding numbers $\mathbf{M}=(M_1,M_2)$ are given respectively by \cite{berry1985}
\begin{equation}\label{eq:RB_parameters}
\begin{split}
&A_{\mathbf{M}}^2 = m^2 a^2 b^2 / \pi^3 \ep T_{\mathbf{M}} \\
&T_{\mathbf{M}} = [2m(M_1^2 a^2 + M_2^2 b^2)/\ep]^{1/2} \\
&S_{\mathbf{M}} = 2\ep T_{\mathbf{M}} .
\end{split}
\end{equation}
Consider, for instance, interference terms between $\mathbf{M}=(M_1,M_2)$ and $\mathbf{M}_p=(M_2,M_1)$. Setting $\hbar=1$ for simplicity, the cosine arguments in (\ref{eq:global_variance_off-diag}) have the following form:\cite{wickramasinghe05, ma10mk}
\begin{equation}
\begin{split}
&S_{\pm} = 2\ep (T_{\mathbf{M}} {\pm} T_{\mathbf{M}_p}) \\
&2\ep T_{\mathbf{M}} = 2 [2m ab \ep ( M_1^2 \al^{1/2} +M_2^2 \al^{-1/2})]^{1/2}
\end{split}
\end{equation}
where $\al = a^2 / b^2$ is the aspect ratio of RB. \cite{berry1985} It is then trivially seen that at $\al_{0}=1$ and for $M_1 \neq M_2$, $\partial_\al {S_{-}} \ne{0}$ while $\partial_\al {S_{+}} = 0$.

Parametric averaging is performed via integration with the Gaussian distribution function $\rho(\al)$ whose width is $\ll{1}$, centered at $\al_0=1$. \cite{ma10mk} Clearly, for a square, $\mathbf{M}$ and {$\mathbf{M}_p$} represent the same orbit per a $90^{\circ}$ rotation; for RB with aspect ratios close to unity, that is a near square shape, we observe interference from geometrically similar orbits with nearly equal lengths. We emphasize that since it is the near equality of lengths that matters, this argument can be easily extended to an arbitrary aspect ratio.

\section{Numerical Results}

In what follows, we express all energies in units of the mean level spacing $\De$ by setting $\De=1$.

\subsection{Global Level Number Variance}

\begin{figure}[htp]
\centering
\includegraphics[width=0.4\textwidth]{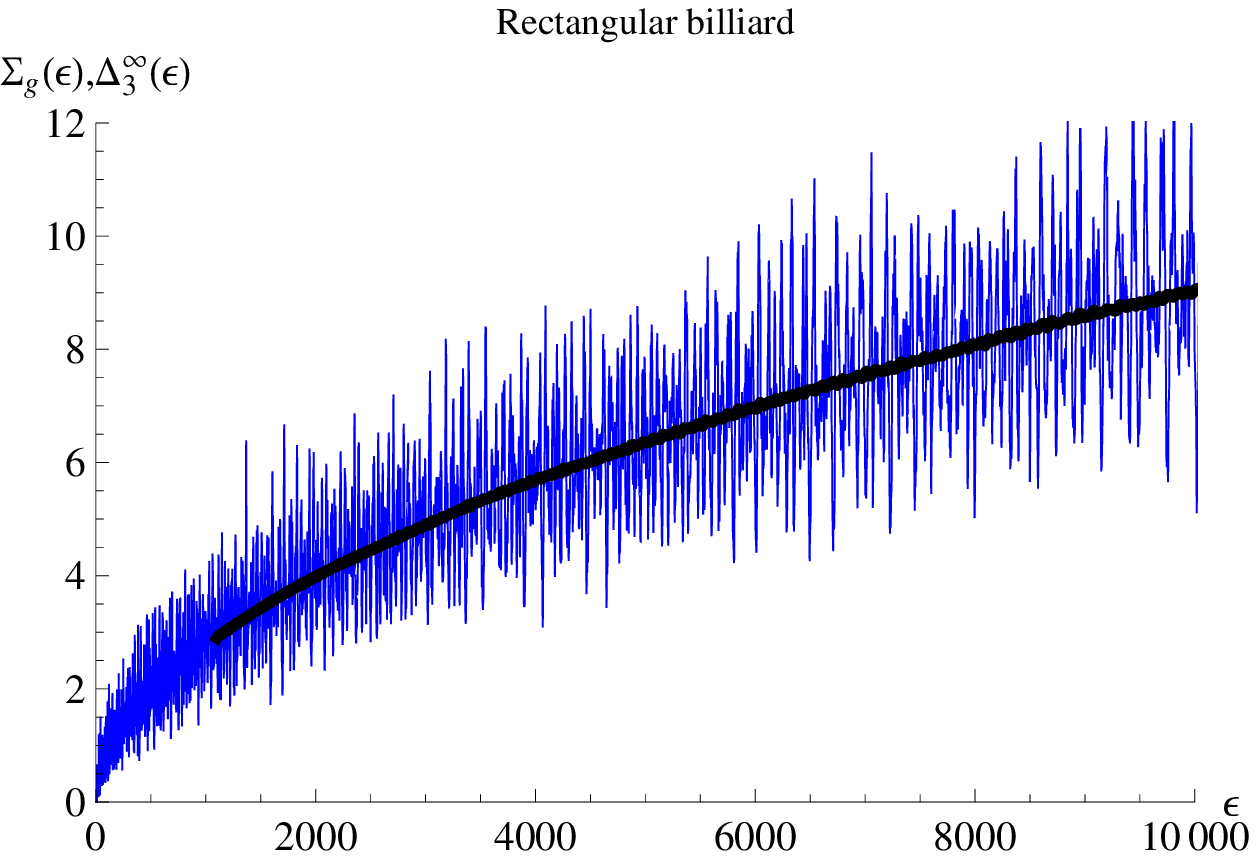}
\includegraphics[width=0.4\textwidth]{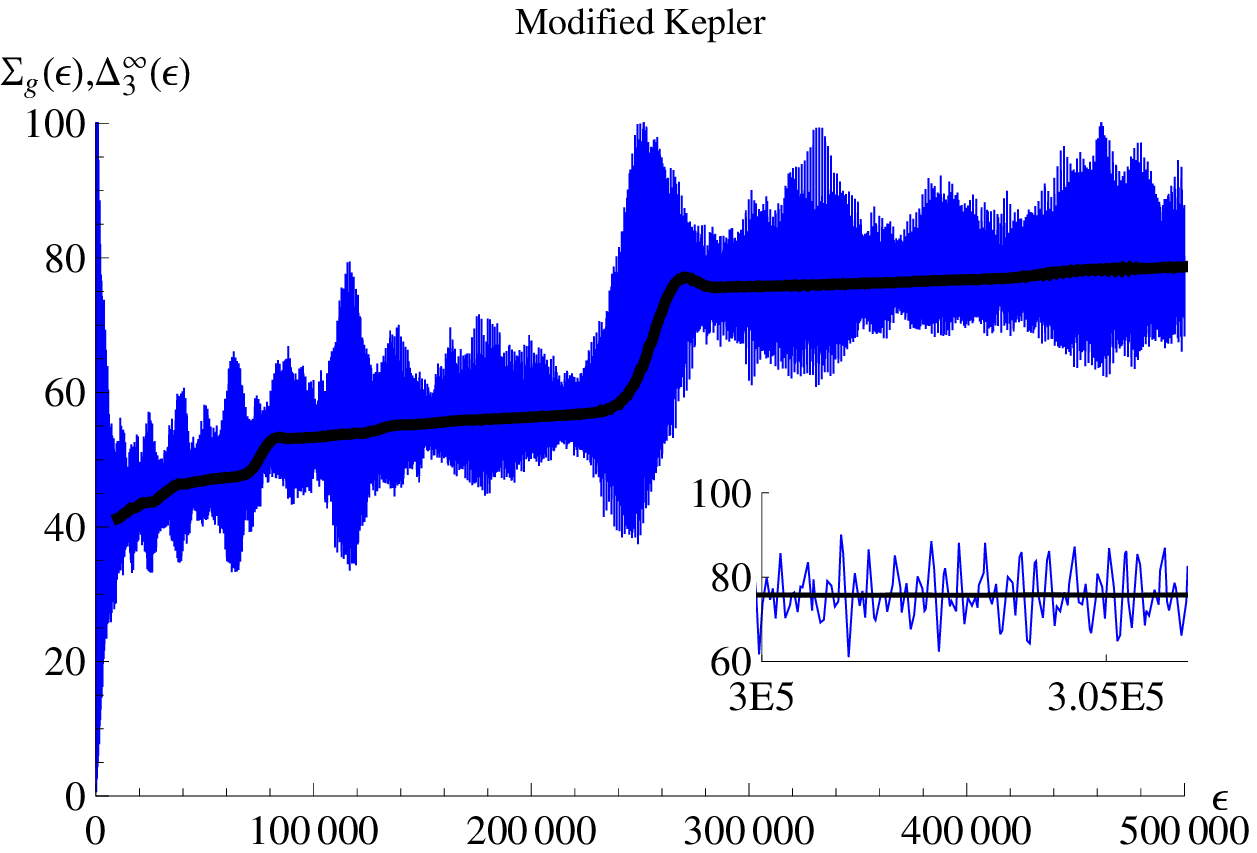} \\
\includegraphics[width=0.4\textwidth]{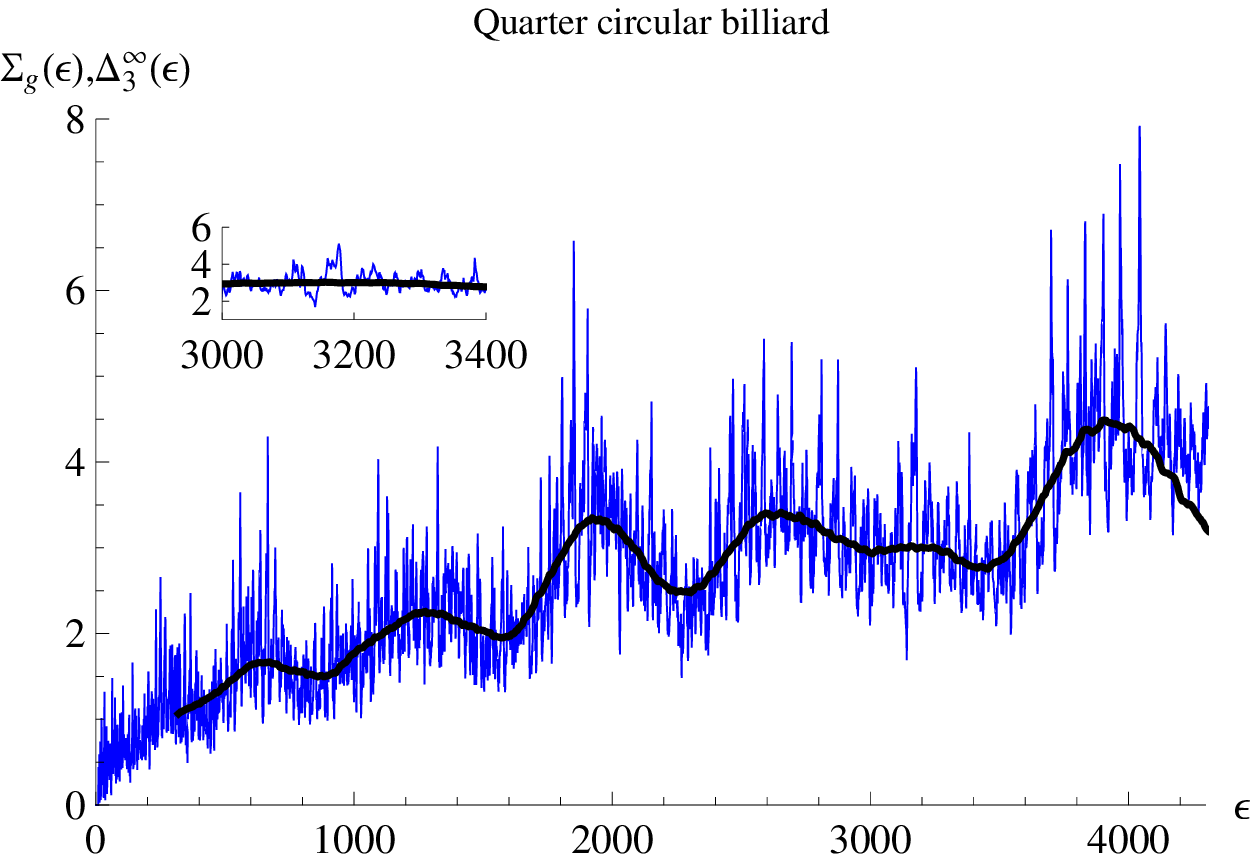}
\includegraphics[width=0.4\textwidth]{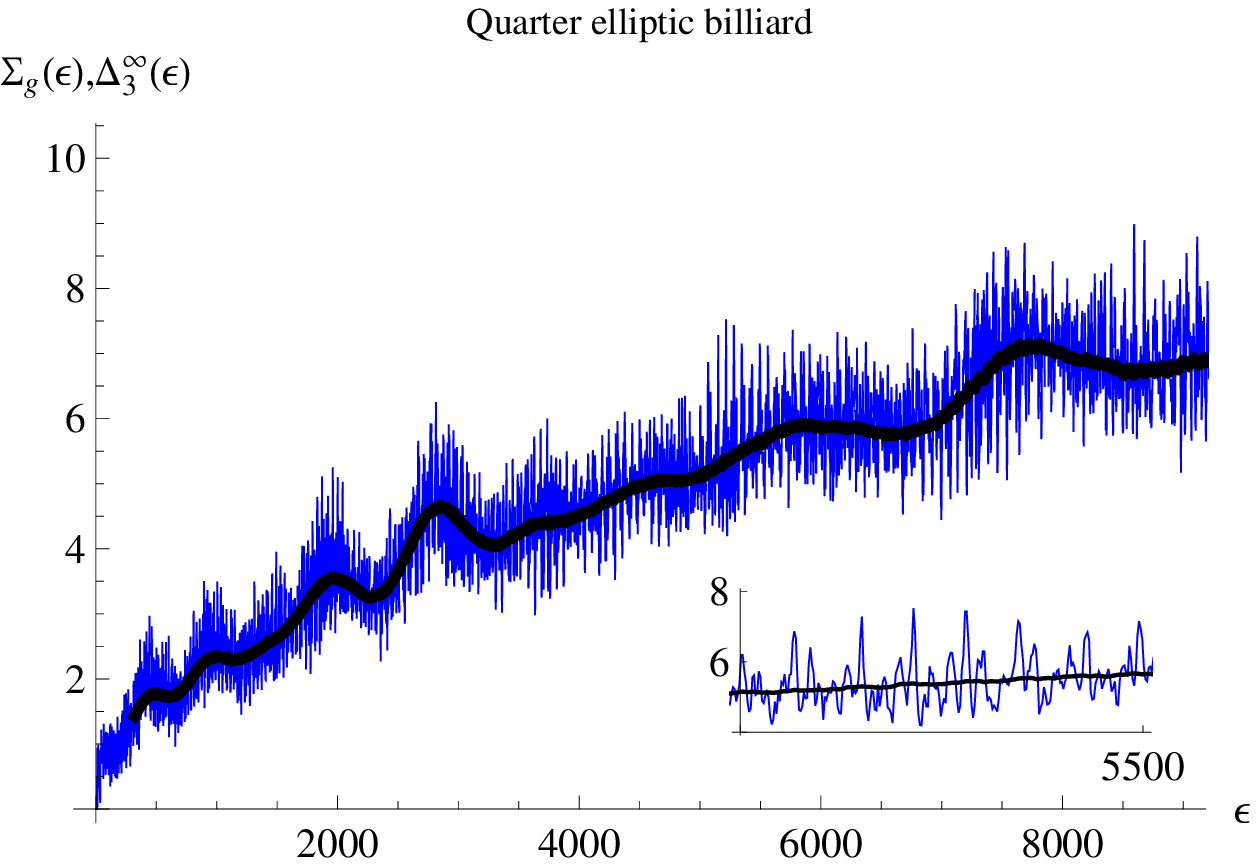} \\
\caption{\label{fig:globalVariance} Comparison between global level number variance and saturation spectral rigidity. Blue line: $\Sigma_g(\ep)$ of rectangular billiard, modified Kepler problem, quarter circular billiard, and quarter elliptic billiard. Black line: $\Delta_3^\infty(\ep)$.  }
\end{figure}

In Fig. \ref{fig:globalVariance}, we plot $\Si_g(\ep)$ vis-a-vis $\De_3^\infty(\ep)$ \footnote[2]{Numerical evaluation of $\De_3(\ep)$ is performed using its definition, Eq. (6) of Ref. \cite{berry1985}.} for RB, MK, CB and EB respectively. \footnote{since it is impossible to perform parametric/ensemble averaging in CB, it is approximated by EB with aspect ratios close to 1.} We observe that $\De_3^\infty(\ep)\sim \sqrt{\ep}$ in RB. \cite{wickramasinghe05, berry1985} In MK, the saturation spectral rigidity exhibits quantum jumps to higher plateaus, while it experiences an overall growth as $\De_3^\infty(\ep)\sim \ep^{1/3}$. \cite{wickramasinghe05, ma10mk} In CB and EB, while scaling overall as $\De_3^\infty(\ep)\sim \sqrt{\ep}$, as expected in a hard-wall billiard, the rigidity exhibits a far more complex behavior than in RB. \cite{ma11eb} While not fully understood, we speculate that its origin may lie in the coherent effects of type-R orbits \cite{waalkens97} of approximately equal  length -- or length multiples -- giving rise to global fluctuations of the level density. \cite{ma11eb} \footnote[3]{Preliminary results indicate that it may be another effect that requires account of interference between POs outside the range of applicability of DA.} As was already mentioned above, persistent oscillations of $\Si_g(\ep)$ around $\Delta_3$ observed in Fig. \ref{fig:globalVariance} cannot be explained in the DA framework. Below, we concentrate on RB in order to explicate the nature of these oscillations.

\onecolumngrid

\twocolumngrid

\subsection{Rectangular Billiard}

\subsubsection{Spectral Staircase}

\begin{figure}
\centering
\subfloat[]{\includegraphics[width=0.4\textwidth]{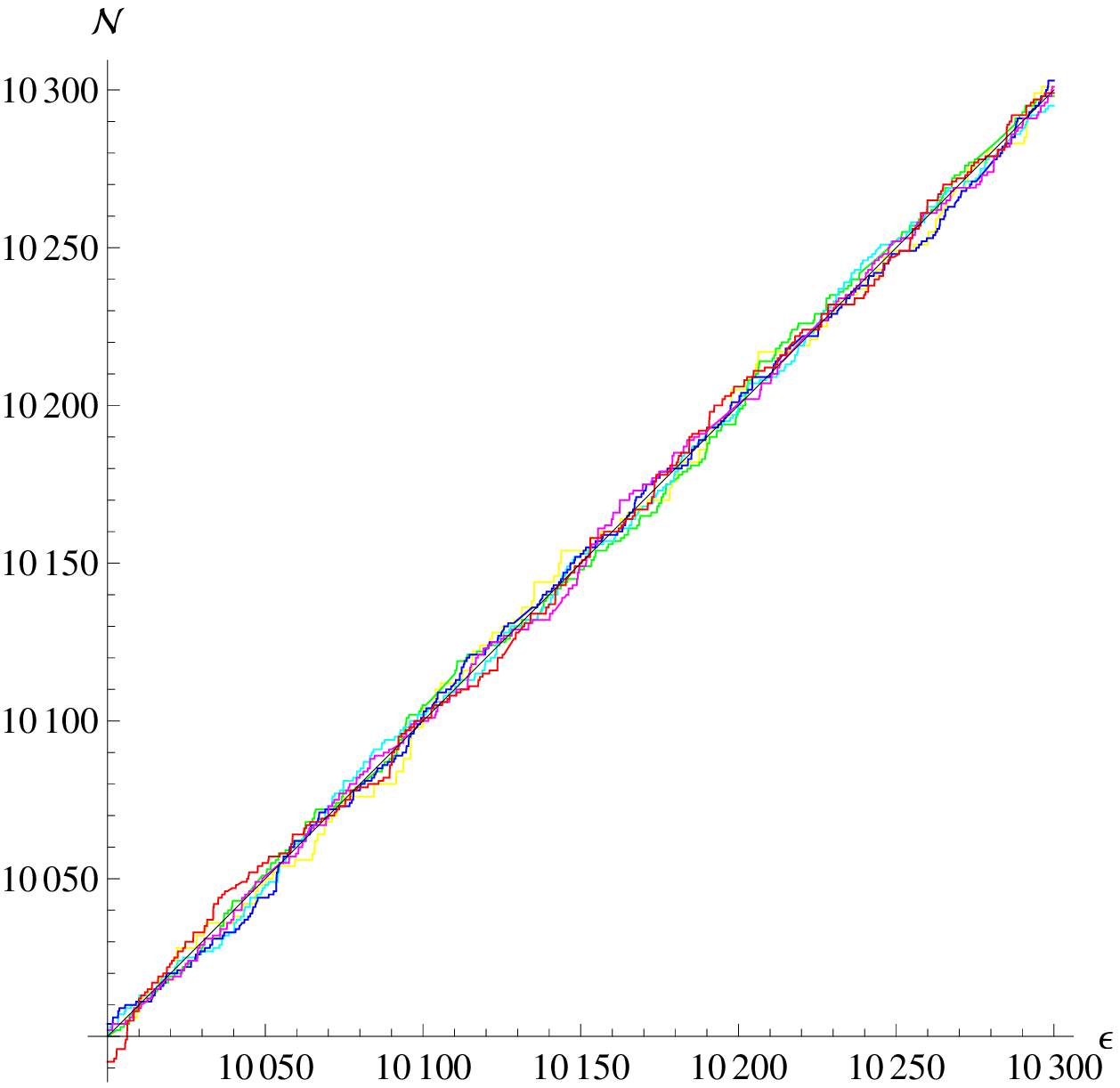}}
\subfloat[]{\includegraphics[width=0.4\textwidth]{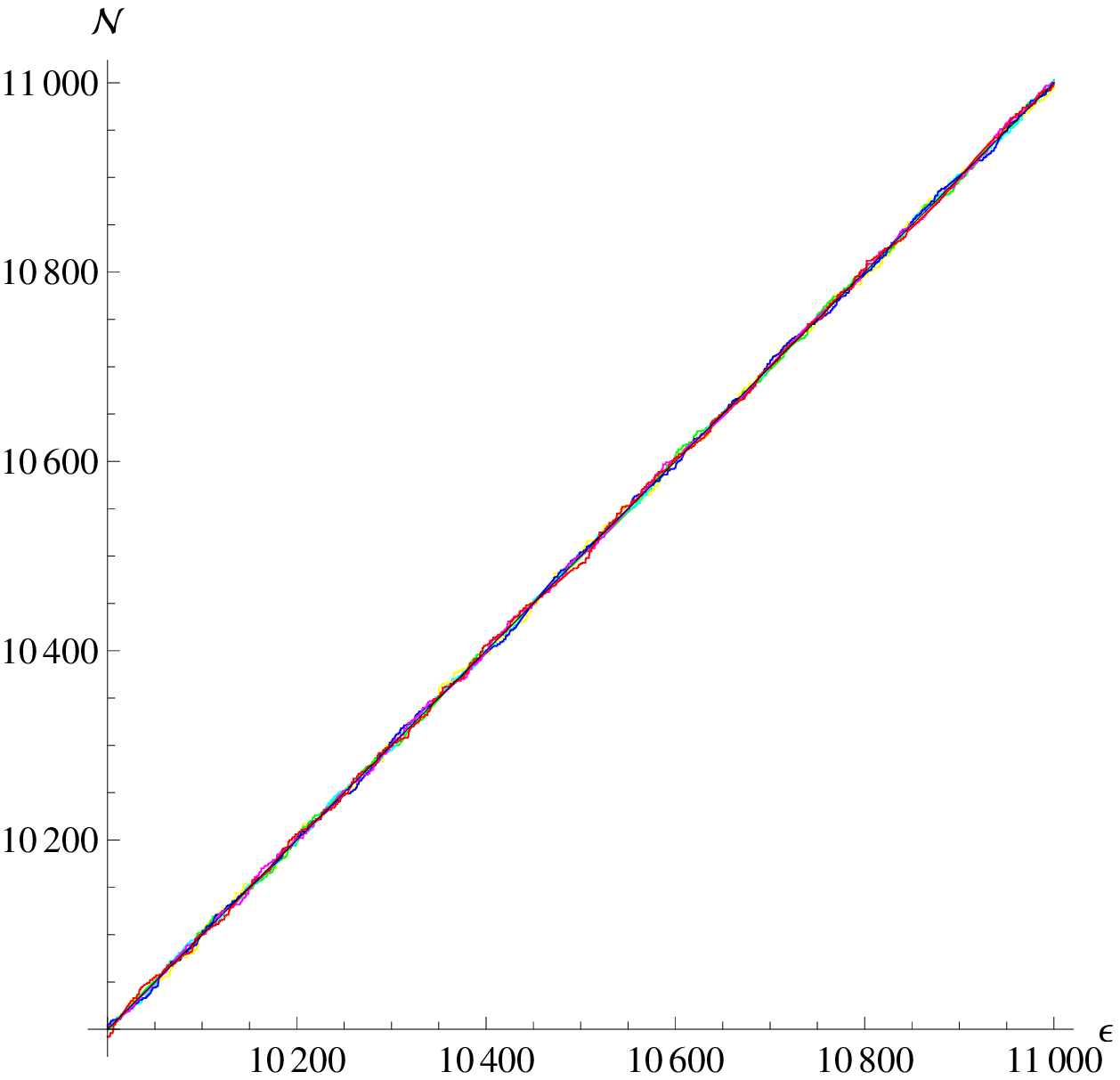}}\\
\subfloat[]{\includegraphics[width=0.33\textwidth]{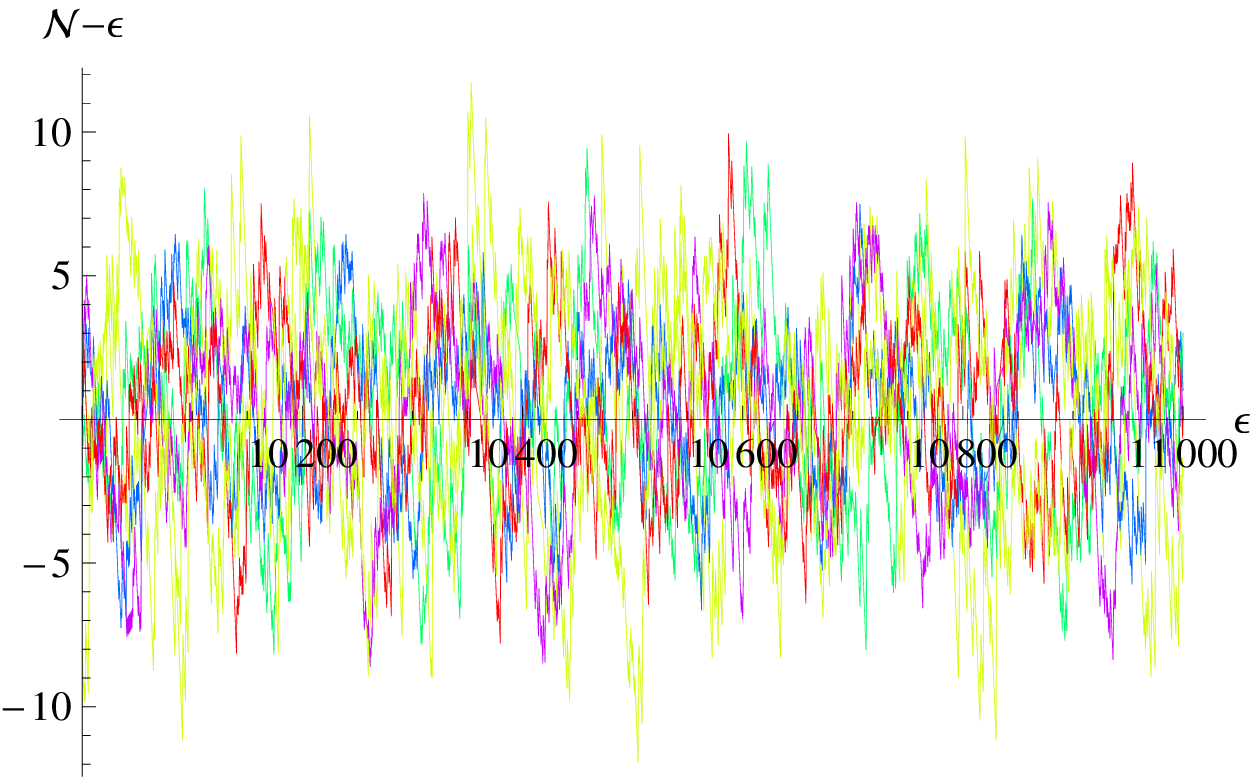}}
\subfloat[]{\includegraphics[width=0.33\textwidth]{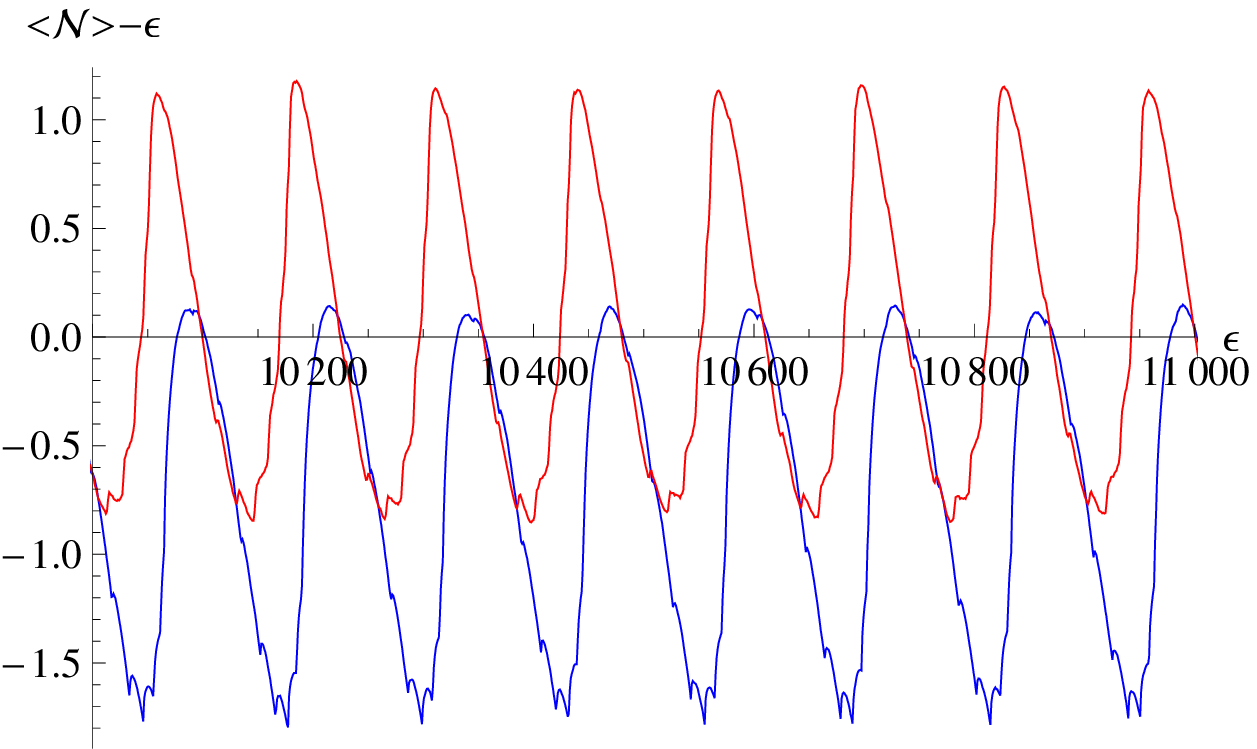}}
\subfloat[]{\includegraphics[width=0.33\textwidth]{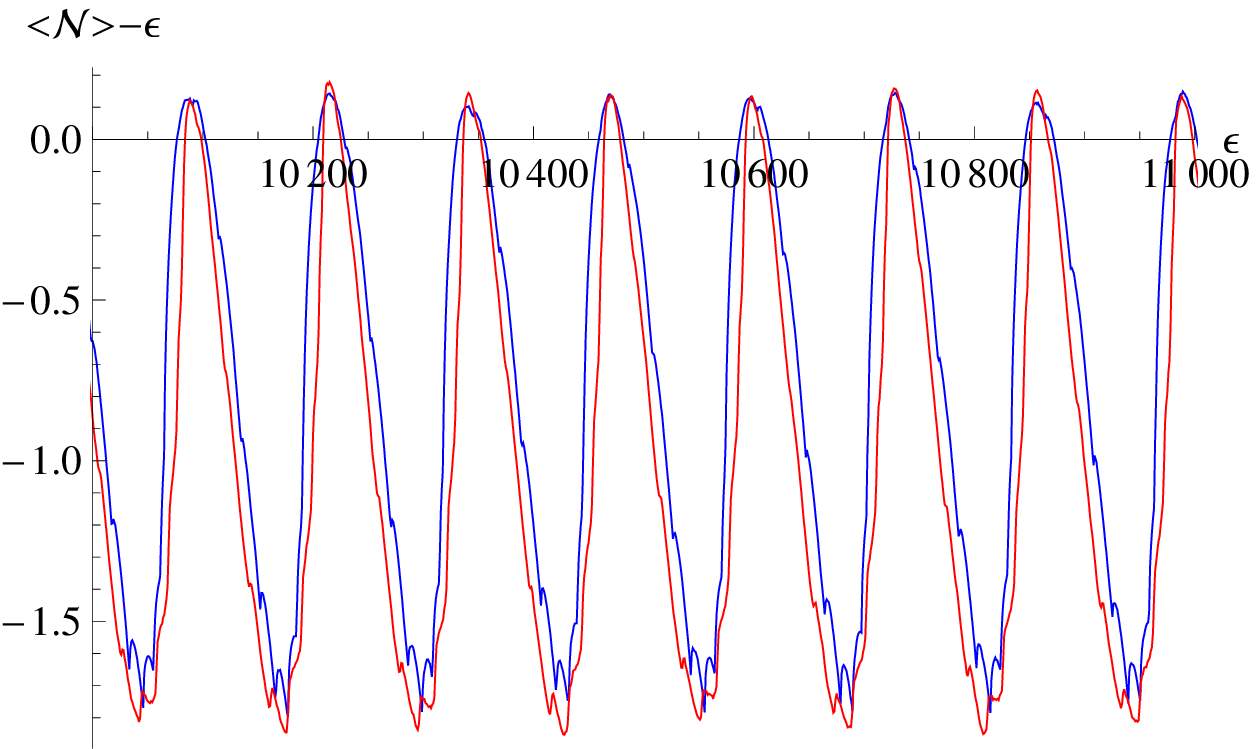}}\\
\caption{\label{fig:deN_E} Rectangular Billiard: (a-b) Spectral staircase for six aspect ratio $\al$'s. Different colors encode different aspect ratios. (c) $\scr{N}(\ep)-\ep$ for six $\al$'s. (d-e) Blue line: $\scr{N}(\ep) - \ep$ calculated by averaging over $10^5$ $\al$'s. Red line: theoretical $\de\scr{N}(\ep)$ calculated from (\ref{eq:deN}) and averaged over $\al$'s and in (e) the theoretical line is shifted 1 downward and 30 rightward. }
\end{figure}

In Figs. \ref{fig:deN_E}a-\ref{fig:deN_E}b, SS is shown respectively over a shorter and longer energy scales  for several values of aspect ratio $\al$. The former reveals noticeable oscillations around the $45^{\circ}$ straight line. To further emphasize this point, in Fig. \ref{fig:deN_E}c, we plot $\scr{N}(\ep) - \ep$ for these $\al$'s. It is natural to anticipate that upon $\al$-averaging, $\scr{N}(\ep) - \ep$ and the theoretical evaluation of $\delta\scr{N}(\ep)$ using (\ref{eq:deN}) should vanish. However, numerical simulation shows neither to be the case, as seen from Fig. \ref{fig:deN_E}d. While a translation of the latter downward and rightward bring the two into congruence, as seen from Fig. \ref{fig:deN_E}e, we do not fully understand the nature of this phenomenon \footnote{Per Eq. (A17) in Ref. \cite{wickramasinghe05}, we can only account for 1/4 downward translation.}. Clearly, it is inherent to the nature of parametric averaging and of the PO theory and is not an artifact of the numerical calculation. We remark that the proximity of $\langle\scr{N}(\ep)\rangle-\ep$ to 0 can be seen as a measure of performance of parametric averaging in attaining ensemble averaging. Comparing its magnitude to that of $\scr{N}(\ep)-\ep$ indicates that it does quite well.

\subsubsection{Global Level Number Variance and Correlation Function of Spectral Staircase}

\begin{figure}[htp]
\subfloat[]{\includegraphics[width=0.4\textwidth]{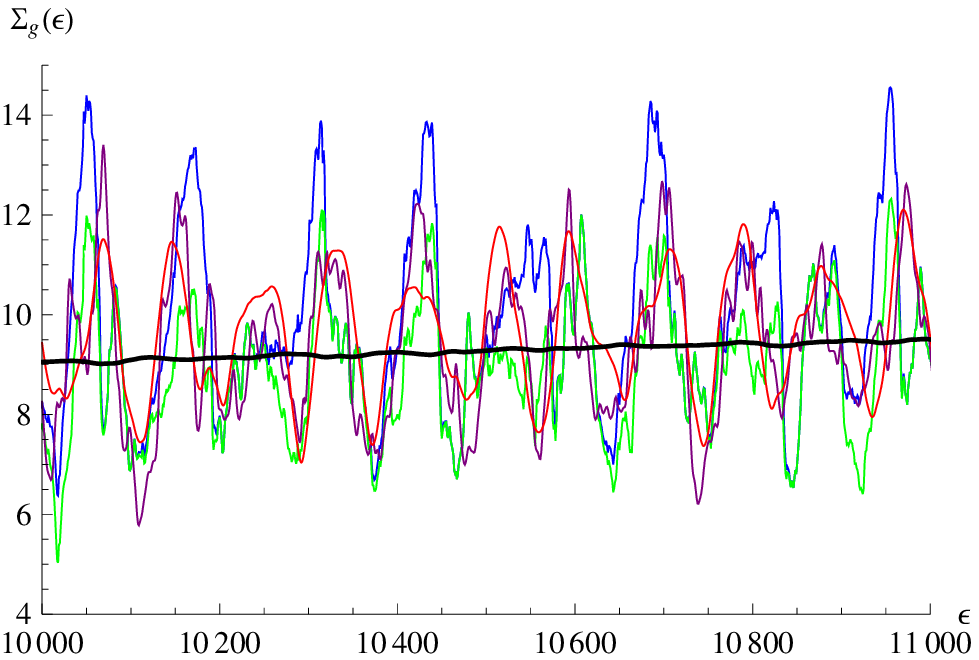}}
\hspace{24pt}
\subfloat[]{\includegraphics[width=0.4\textwidth]{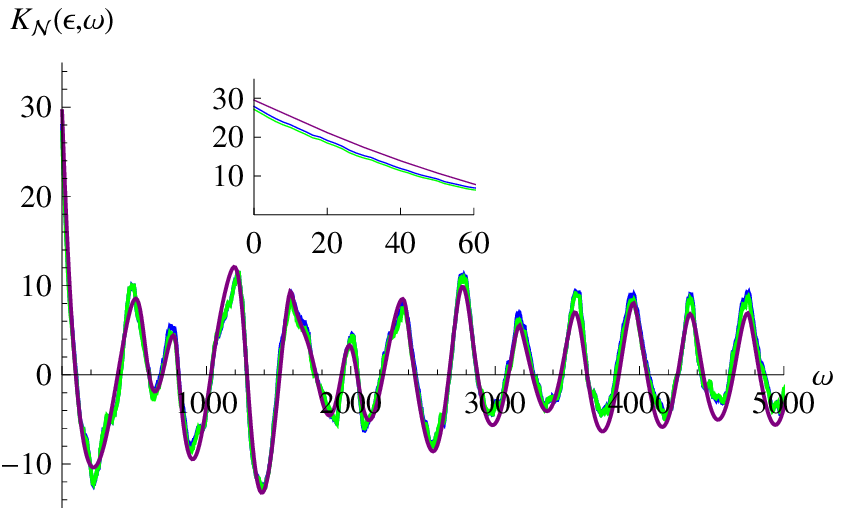} }\\
\caption{\label{fig:KN} Rectangular billiard: (a) Comparison between numerical and theoretical $\Si_g$. Black line: saturation spectral rigidity. Blue line: numerical $\Si_g$ calculated from $\langle [\scr{N}-\ep]^2\rangle$. Green line: numerical $\Si_g$ calculated from $\langle [\scr{N}- \langle \scr{N} \rangle ]^2\rangle$.
Purple line: theoretical $\Si_g$ calculated from (\ref{eq:global_variance_full}) after averaging over aspect ratios. Red line: theoretical $\Si_g$ calculated from diagonal approximation plus interference between terms $(M_1, M_2)$ and $(M_2, M_1)$ with $M_1\neq M_2$. (b) Correlation function of spectral staircase with $\ep=10^5$; insert shows small $\om$ behavior. Blue line: calculated from $\langle (\scr{N}(\ep_1)-\ep_1) (\scr{N}(\ep_2)-\ep_2) \rangle$. Green line: calculated from $\langle (\scr{N}(\ep_1)-\langle \scr{N}(\ep_1)\rangle) (\scr{N}(\ep_2)-\langle \scr{N}(\ep_2)\rangle) \rangle$. Purple: theory with diagonal approximation and parametric} averaging. 
\end{figure}

In Fig. \ref{fig:KN}a, we plot $\Si_g(\ep)$. Theoretical fit of the numerical data is quite good -- given the limitations discussed above -- and underscores importance of the non-diagonal terms.

In Fig. \ref{fig:KN}b, we plot $K_\scr{N}(\ep,\om)$. Theoretical and numerical curves are in excellent agreement, which proves applicability of DA (\ref{eq:Kdiag}). We specifically point out the small $\om$ behavior in the insert of Fig. \ref{fig:KN}b, which is described by the small $\om$ expansion in (\ref{eq:Kdiag}) and corresponds to the $\delta(\om)$ term in the level density correlation function. \cite{wickramasinghe05,wickramasinghe08}

\section{Summary}

We examined the global level number variance and the correlation function of spectral staircase in generic integrable systems with no extra degeneracies. We demonstrated that the global level number variance exhibits persistent oscillations around the saturation spectral rigidity. These oscillations cannot be explained in the diagonal approximation framework and require an account of interference terms.

Conversely, the correlation function of spectral staircase is well explained by the diagonal approximation. The latter points to the subtlety of the $\om \to 0$ limit since mathematically interference is destroyed by a finite $\om$. 

In the future, we need to gain greater insight into integrable systems beyond the better understood rectangular billiards. For instance, we need to develop a quantitative description of the interference effects leading to the oscillations of the global level number variance in the modified Kepler problem. Larger-scale oscillations of the saturation spectral rigidity in circular and elliptic billiards, which also appear to be a product of the periodic orbit interference and are thus outside of the range of applicability of the diagonal approximation framework, are of great interest and call for further investigation. Properties of the spectral staircase, as well as its description using parametric averaging, require a closer examination as well.

\begin{acknowledgements}
Computational part of this work was supported by the Ohio Supercomputer Center.
\end{acknowledgements}

\bibliographystyle{plain}

\onecolumngrid

\onecolumngrid

\twocolumngrid

\end{document}